\documentclass[conference]{IEEEtran}

\ifCLASSINFOpdf
\else
\fi

\usepackage{amssymb}
\usepackage{amsmath}
\usepackage{booktabs}
\usepackage{mdwlist}

\usepackage{dsfont}
\usepackage{amssymb}
\usepackage{amsmath}
\usepackage{booktabs}
\usepackage{mdwlist}
\usepackage{caption}
\usepackage{color}
\usepackage{algorithm}
\usepackage{algorithmic}
\usepackage{threeparttable}
\usepackage{multirow}
\usepackage{color}
\usepackage{graphicx}
\usepackage{subfigure}
\usepackage{latexsym}
\usepackage{supertabular,booktabs}
\usepackage{leftidx}

\begin{document}
%
% paper title
% Titles are generally capitalized except for words such as a, an, and, as,
% at, but, by, for, in, nor, of, on, or, the, to and up, which are usually
% not capitalized unless they are the first or last word of the title.
% Linebreaks \\ can be used within to get better formatting as desired.
% Do not put math or special symbols in the title.
\title{An Efficient Method to Transform SAT problems to Binary Integer Linear Programming Problem}

% author names and affiliations
% use a multiple column layout for up to three different
% affiliations
%\author{\IEEEauthorblockN{Michael Shell}
%\IEEEauthorblockA{School of Electrical and\\Computer Engineering\\
%Georgia Institute of Technology\\
%Atlanta, Georgia 30332--0250\\
%Email: http://www.michaelshell.org/contact.html}
%\and
%\IEEEauthorblockN{Homer Simpson}
%\IEEEauthorblockA{Twentieth Century Fox\\
%Springfield, USA\\
%Email: homer@thesimpsons.com}
%\and
%\IEEEauthorblockN{James Kirk\\ and Montgomery Scott}
%\IEEEauthorblockA{Starfleet Academy\\
%San Francisco, California 96678--2391\\
%Telephone: (800) 555--1212\\
%Fax: (888) 555--1212}}

% conference papers do not typically use \thanks and this command
% is locked out in conference mode. If really needed, such as for
% the acknowledgment of grants, issue a \IEEEoverridecommandlockouts
% after \documentclass

% for over three affiliations, or if they all won't fit within the width
% of the page, use this alternative format:
%
\author{\IEEEauthorblockN{Wenxia Guo\IEEEauthorrefmark{1},Jin Wang\IEEEauthorrefmark{1},Majun He\IEEEauthorrefmark{1}, Xiaoqin Ren\IEEEauthorrefmark{1},
Wenhong Tian\IEEEauthorrefmark{1},Qingxian Wang\IEEEauthorrefmark{2}}
\IEEEauthorblockA{\IEEEauthorrefmark{1} School of Information and Software Engineering\\
University of Electronic Science and Technology of China, Chengdu, Sichuan\\
Email: 359873769@qq.com\\
}
\IEEEauthorblockA{\IEEEauthorrefmark{2}
Chongqing Institute of Green and Intelligent Technology, Chinese Academy of Sciences\\
Email:tian\_wenhong@uestc.edu.cn, tianwenhong@cigit.ac.cn\\
}
}

% use for special paper notices
%\IEEEspecialpapernotice{(Invited Paper)}

% make the title area
\maketitle

% As a general rule, do not put math, special symbols or citations
% in the abstract
\begin{abstract}
 In computational complexity theory, a decision problem is NP-complete when it is both in NP and NP-hard. Although a solution to a NP-complete can be verified quickly, there is no known algorithm to solve it in polynomial time. There exists a method to reduce a SAT (Satifiability) problem to Subset Sum Problem (SSP) in the literature, however, it can only be applied to small or medium size problems. Our study is to find an efficient method to transform a SAT problem to a mixed integer linear programming problem in larger size. Observing the feature of variable-clauses constraints in SAT, we apply linear inequality model (LIM) to the problem and propose a method called LIMSAT. The new method can work efficiently for very large size problem with thousands of variables and clauses in SAT tested using up-to-date benchmarks.
\end{abstract}

keywords: SAT(Satisfiability problem); BinaryILP(Integer Linear programing); 3SAT; Reduction

% For peer review papers, you can put extra information on the cover
% page as needed:
% \ifCLASSOPTIONpeerreview
% \begin{center} \bfseries EDICS Category: 3-BBND \end{center}
% \fi
%
% For peerreview papers, this IEEEtran command inserts a page break and
% creates the second title. It will be ignored for other modes.
\IEEEpeerreviewmaketitle

\section{Introduction}
% no \IEEEPARstart
P problems are the class of problems that can be solved in polynomial time. This means they are problems which are solvable in time $O(n^k)$ in Big $O$ notation for any constant $k$, where $n$ is the size of the input of the problem. NP problems are the set of problems that can be verified in polynomial time as a function of the given input size \cite{Vanoye2011Survey} using a nondeterministic Turing machine. This means if there is a ¡°certificate¡± of a solution, then the ¡°certificate¡± can be proved to be correct in time polynomial in the size of the input to the problem \cite{Cormen2005Introduction}. In 1971, Cook \cite{Cook1971The} defined that problem X polynomial reducible to problem Y if arbitrary instances of problem X can be solved using polynomial number of standard computational steps, plus polynomial number of calls to oracle that solves problem Y. Therefore, the definition of NP complete is a problem Y in NP with the property that for every problem X in NP, $X \leq\leftidx{_p}Y,$ that is problem X can be polynomial-time reducible to problem Y. NP-complete problems constitute the class of the most difficult possible NP problems \cite{Vanoye2011Survey}.

NP-complete problems can be divided into six basic genres \cite{karp1972Reducibility}, i.e., packing problems, covering problems, constraint satisfaction problems, sequencing problems, partitioning problems, numerical problems.

Constraint satisfaction problems include Circuit Satisfiability problems, Satisfiability problem, 3SAT. A specific situation of SAT is 3SAT that each clause of it has exact three literals, which correspond to distinct variables or the negative form of these variables. The computational complexity of SAT problem in the worst case is $O(2^n)$, where n is the number of variables. Because SAT can be transformed to 3SAT, it has similar computational complexity as SAT.

The question whether an arbitrary Boolean formula is satisfiable cannot be solved within polynomial time. A formula with n variables, possibilities of variables assignment can reach to $2^n$. If formula length ${\O}$ is a polynomial length about n, then it will take $¦¸(2^n)$ for each assignment. It is a superpolynomial length about to formula length. Due to this fact, this paper aims to propose an efficient method to transform SAT problems to a mixed integer linear programming problem to reduce the handling time of SAT problem.

\subsection{Related Work}
Conflict-driven clause learning (CDCL) is an efficient method for solving Boolean satisfiability problems (SAT). Up to now, many heuristics are added in it to improve performance, for example, restart, Variable State Independent Decaying Sum (VSIDS). Hidetomo et al. \cite{Nabeshima2017Coverage} focus on clause reduction heuristic, which aims to suppress memory consumption and sustain propagation speed. In their study, the reduction consists of two parts: evaluation criteria and reduction strategy. The first step measuring the usefulness of learnt clauses is using LBD (literals blocks distance) and the latter one using a new strategy based on the coverage of used LBDs is to select removing clauses according to the criteria. In experiments, they compare Glucose schema and Coverage schema. The result shows that the new strategy improves performance for both SAT and UNSAT instances. Another strategy used to improve SAT solver is called HSAT (Hint SAT). It is proposed by Jonathan Kalechstain et al. \cite{Kalechstain2015Hints} to cut the searching space by using a hint-based partial resolution-graph to get a solution faster. For hint generation, they chiefly use two heuristics. The first one is Avoiding Failing Branches (AFB) which avoids the solver spending too much time on explored branches which are made of decision variables. The second heuristic is Random Hints (RH) which aims to create hints that contradict the instance. This algorithm is based on random assignments and satisfiability checking. Experiments show that AFB can solve 113 instances from SAT 2013 within half an hour where the total number of satisfiability instances is 150. The experimental consequents of SAT 2014 are almost the same.

Gilles Audemard et al. \cite{Audemard2016} study how to measure SAT instances. They give 5 indicators: the number of decision levels, the number of decisions between two conflicts, the number of successive conflicts, the number of non-binary glue clauses and the number of unit propagation. They also mention the restart polarity policy which is added to Glucose. The new version of Glucose solves 20$\%$ more problems than the original one and increases the speed for UNSAT instances. Further, Mathan Mull et al. \cite{Mull2016On} analysis the structure of industrial benchmarks. Previous studies hold that the reason why CDCL solver is efficient for industrial benchmarks is due to its ¡°good community structure¡± (high modularity). However, Mathan Mull et al. get the different result. They use random unsatisfiable instances produced by ¡°pseudo-industrial¡± community attachment model to do experiment. The result shows that community structure is not adequate to explain the good performance of CDCL on industrial benchmarks.

Symmetry is another characteristic of SAT problem. Jo Devriendt et al \cite{Devriendt2017Symmetric} present symmetric explanation learning (SEL), in which symmetric clauses are learned only when they are unit or conflicting. 1300 benchmark instances indicate that among GLUCOSE, BREAKID, SEL, SP, SLS, SEL outperforms other four solver configurations and as a dynamic symmetry handling technique, SEL is the first one competitive with static symmetry breaking which is known to the most effective way to handle symmetry. In terms of symmetry, C.K Cuong et al. \cite{Cuong2016Computing} as well propose a method transforming unavoidable sub-graphs to SAT, which can make up for the shortcomings of SAT solvers.

With the development of Machine Learning, combing SAT problem with Machine Learning is a good idea. Quanrun Fan et al. \cite{Fan2015Clustering} use clustering method basing on divide and conquer to deal with Boolean satisfiability problems. In this way, the original problem will be divided into many small ones. Therefore, the overall runtime reduces. In this algorithm, clauses are clustered into a group according to the similarity between clauses. They use Stoer-Wagner algorithm the minimum cut of undirected graph to partition clauses. Thus, the number of variables preventing clause group partition (cut variables) is down, and eliminating these variables will be easier.

In SAT Competition 2016 \cite{SAT2016}, the best solver for main-Crafted benchmarks is tc\_glucose, which solves 58 instances in total. It combines CHBR\_glucose with tb\_glucose. There are two techniques: Variable State Independent Decaying Sum(VSIDS) decision heuristic \cite{Matthew W2005Chaff} and conflict history-based branching heuristic(CHB)\cite{JiaHuiLiang2016Exponential}. The first one is good at dealing with big problem while the latter works well with small problem. So if the number of variables is under 15000, CHBR\_glucose uses CHB. Because of the fact that once a variable scored by VSIDS, ties happen frequently. To avoid this, tb\_glucose updates VSIDS sores after getting learned clauses and computes 1/(LBD of a clause) for each variable in that clause\cite{Tom¨¢ Balyo2016Proceeding}. This is called TBVSIDS. In tc\_glucose, TBVSIDS is a default decision heuristic and CHB is activated when the number of variables is under 15000.

\section{The method}

\subsection{Reduction from 3SAT to SSP}
The Subset-Sum problem (SSP) is the problem to find a subset of numbers which add up to a target value from a given set of numbers. The approach of transforming 3SAT to SSP is introduced in \cite{Cormen2005Introduction}, for completeness, we restate the approach in the following.

Given a 3-CNF formula with $n$ variables and $k$ clauses. The most significant digits are labeled by $n$ variables, and the least significant digits are labeled by $k$ clauses. The construction is as follows:
  \begin{enumerate}
   \item The target t has a 1 in each digit labeled by a variable and a 4 in each digit labeled by a clause.
   \item For each variable $x_i$, set $S$ contains two integers $v_i$ and $v'_i$. Each $v_i$ and $v'_i$ equals to 1 in the digit labled by $x_i$ and equals to 0 in other digits. If $C_j$ contains $x_i$, then the digit labeled by $C_j$ in vi has a 1. If $©´x_i$ appears in $C_j$, then the digit labeled by $C_j$ in $v'_i$ has a 1.
   \item For each clause $C_j$, set S contains two integers $s_j$ and $s'_j$. Except the digit labeled by $C_j$, other digits all are 0. $s_j$ has a 1 in the digit labeled by $C_j$, $s'_j$ has a 2 in the same digit. $s_j$ and $s'_j$ are ¡°slack variables¡±, which help us to get each clause-labeled digit to sum to 4 which is the target value.
   Table. \ref{Reduction} shows an example of reduction from 3SAT to SSP (providing the example of 3 variables and 7 clauses)

   Example: 3 variables, 4 clauses
   \begin{center}
   $\phi$ \ = \ $C_1$ \ $\wedge$ \ $C_2$ \ $\wedge$ \ $C_3$ \ $\wedge$ \ $C_4$ \qquad (1)
   \end{center}
   within the formula, the four clauses are : \\
   $C_1$  = ($x_1$  $\vee$ ${\neg x_2}$ $\vee$${\neg x_3}$), \ $C_2$  =  (${\neg x_1}$  $\vee$  ${\neg x_2}$  $\vee$ ${\neg x_3}$), \\
   $C_3$  = ($x_1$  $\vee$ ${\neg x_2}$ $\vee$  $x_3$), \ $C_4$  = ($x_1$  $\vee$ $x_2$  $\vee$  $x_3$)

   The target $t$ in the SSP is 1114444, so the job is to find a subset which has a sum equals to the target $t$. By applying the dynamic programming, it is possible to solve this SSP problem.
One can see that there are totally $(n+k)$-digit for each number in the final SSP problem with total $2(n+k)$ numbers. So this approach can work only for small or medium size problem.
  \end{enumerate}
\begin{table}[!hbp]
\caption{Reduction from 3SAT to SubSet-Sum} \label{Reduction}
\begin{center}
\begin{tabular}{|c|c|c|c|c|c|c|c|c|}
\hline
 & $X_1$ & $X_2$ & $X_3$ & $C_1$& $C_2$& $C_3$& $C_4$ \\
\hline
 $v_1$  & 1 & 0 & 0 & 1 & 0 & 0& 1 \\
\hline
$v'_1$  & 1 & 0 & 0 & 0 & 1 & 1& 0 \\
\hline
$v_2$   & 0 & 1 & 0 & 0 & 0 & 0& 1 \\
\hline
$v'_2$  & 0 &1 & 0 & 1 & 1 & 1& 0 \\
\hline
$v_3$   & 0 & 0 & 1 & 0 & 0 & 1& 1 \\
\hline
$v'_3$  & 0 & 0 & 1 & 1 & 1 & 0& 0 \\
\hline
$s_1$   & 0 & 0 & 0 & 1 & 0 & 0& 0 \\
\hline
$s'_1$  & 0 & 0 & 0 & 2 & 0 & 0& 0 \\
\hline
$s_2$   & 0 & 0 & 0 & 0 & 1 & 0& 0 \\
\hline
$s'_2$  & 0 & 0 & 0 & 0 & 2 & 0& 0\\
\hline
$s_3$   & 0 & 0 & 0 & 0 & 0 & 1& 0 \\
\hline
$s'_3$  & 0 & 0 & 0 & 0 & 2& 0& 0\\
\hline
$s_4$   & 0 & 0 & 0 & 0 & 0 & 0& 1 \\
\hline
$s'_4$  & 0 & 0 & 0 & 0& 0 & 0& 2\\
\hline
$ t $   & 1 & 1 & 1 & 4 & 4 & 4 & 4\\
\hline
\end{tabular}
\end{center}
\end{table}

\subsection{Transform from SAT to 0-1 Integer Linear programing Model}

Observing the difficulty to handle the large number in the approach of reduction 3SAT to SSP, we consider a new approach to solve SAT. Instead of treating each number independently in the reduction of 3SAT to SSP, we treat each bit as an element in the matrix of SSP. So by setting the matrix in SSP as $A$, the integer linear programming (ILP) formulation for transforming SSP to 0-1 ILP becomes to $x A=b$, where $b$ is the target value and $x$ is the solution we are looking for. To meet the constraints of SAT, we can build a linear inequalities model as follows:
\begin{enumerate}
  \item Preparing a SAT problem in CNF format;
  \item Reduction from a SAT problem to SSP by the method introduced in \cite{Cormen2005Introduction}; obtaining a matrix $A$ which has $2(n+k)$ by $(n+k)$ dimension, we denote the matrix $A_1$ which is $2n$ by $n$ from A, and $A_2$ which is $2n$ by $k (n+1:n+k)$ from $A$.
  \item To meet the constraints, we need  $x{A_1} \leq b_1$, $b_2 \leq x{A_2} \leq b_3$ , where $b_1$ and $b_2$ are all ones, and $b_3$ are all threes. We only need $b_1$ and $b_2$ for our problem. Set target value as an array in $b$, where $b=[ones(1,n),-ones(1,k)]$, and solve integer linear equation
        $x[A_1,-A_2] \leq b$, where $b=[b_1,-b_2]$, if there exits solution to $x$, then the original SAT problem is satisfiable, otherwise, it is not. Our model is as following:
  \begin{center}
         \par{min \ $c^{T}x$} \qquad (2)
         \par{s.t. \ $xA\leq b$} $x\in(0,1)$
         \leftline{where c is coefficient(default as all ones)}
 \end{center}
\end{enumerate}

\section{Experimental Results}

 The test cases come from SAT 2016 competition \cite{Crafted}. There are 5 categories: Application Benchmarks from Main/Parallel Tracks, Crafted Benchmarks from Main/Parallel Tracks, Agile Track Benchmarks, Random Track Benchmarks and Incremental Track Benchmarks. In our research, we focus on main-crafted Track in which a majority of instances come from Crafted Benchmarks from main Tracks. In fact, these instances are limited in 5000 seconds. We carry out our algorithm in Gurobi 7.5.1. In order to avoid being out of memory, we use sparse matrix as input. All the experiments were performed on Intel xeon CPU(2.4GHz) with 20G memory which is similar to the configuration in SAT 2016 \cite{Crafted}. The time limit was set to 5000s.

 In order to verify the correctness of our algorithm, firstly we test one hundred instances named uf250-1065 coming from \cite{SATLIB}. These instances with 250 variables and 1065 clauses are all SAT, and our experimental result is the same as the given result. Table \ref{Resultuf250} containing part of the testing result shows the algorithm proposing in this paper can solve problems correctly.
\begin{table}[!hbp]
 \caption{Part of result of uf250-1065 instances (others are solved within a few seconds)} \label{Resultuf250}
 \begin{center}
 \begin{tabular}{|c|c|c|c|c|}
 \hline
 Filename & Variable & Clause & Result & Time(s) \\
 \hline
 $uf250-02.$ cnf & 250 & 1065 & sat & 63 \\
 \hline
 $uf250-024.$ cnf & 250 & 1065 & sat & 24 \\
 \hline
 $uf250-029.$ cnf & 250 & 1065 & sat & 74 \\
 \hline
 $uf250-054.$ cnf & 250 & 1065 & sat & 31 \\
 \hline
 $uf250-067.$ cnf & 250 & 1065 & sat & 23 \\
 \hline
 $uf250-071.$ cnf & 250 & 1065 & sat & 53 \\
 \hline
 $uf250-086.$ cnf & 250 & 1065 & sat & 44 \\
 \hline
 $uf250-093.$ cnf & 250 & 1065 & sat & 41\\
 \hline
 \end{tabular}
 \end{center}
\end{table}
 Next, we implement our algorithm for SAT competition problems. For main-crafted instances, we test 104, 68 of them is SAT, the rest is unknown. Table \ref{Result-Craft} lists successfully solved instances.

\begin{center}
\tablefirsthead{%
\hline
\multicolumn{1}{|c}{ Filename} &
\multicolumn{1}{|c}{Result} &
\multicolumn{1}{|c|}{CPUTime(s)} \\
\hline}
\tablehead{%
\hline
\multicolumn{3}{|l|}{\small\sl continued from previous page}\\
\hline
\multicolumn{1}{|c}{ Filename} &
\multicolumn{1}{|c}{Result} &
\multicolumn{1}{|c|}{CPUTime(s)} \\
\hline}
\tabletail{%
\hline
\multicolumn{3}{|r|}{\small\sl continued on next page}\\
\hline}
\tablelasttail{\hline}
\topcaption{Successfully solved main-Craft benchmark instances}
\label{Result-Craft}
\begin{supertabular}{|c|c|c|}
craft\_fixedbandwidth-eq-31 & unsat & $0.99$ s \\
 \hline
 craft\_fixedbandwidth-eq-32 & unsat & $0.74$ s \\
 \hline
 craft\_fixedbandwidth-eq-33 & unsat & $0.71$ s \\
 \hline
 craft\_fixedbandwidth-eq-34 & unsat & $0.74$ s \\
 \hline
 craft\_fixedbandwidth-eq-35 & unsat & $0.71$ s \\
 \hline
 craft\_fixedbandwidth-eq-36 & unsat & $1.58$ s \\
 \hline
 craft\_fixedbandwidth-eq-37 & unsat & $1.70$ s \\
 \hline
 craft\_fixedbandwidth-eq-39 & unsat & $0.87$ s \\
 \hline
 craft\_fixedbandwidth-eq-40 & unsat & $0.81$ s \\
 \hline
 craft\_fixedbandwidth-eq-42 & unsat & $0.79$ s \\
 \hline
 craft\_rphp4\_065 & unsat & $107.51$ s \\
 \hline
 craft\_rphp4\_070 & unsat & $233.01$ s \\
 \hline
 craft\_rphp4\_075 & unsat & $209.19$ s \\
 \hline
 craft\_rphp4\_080 & unsat & $136.75$ s \\
 \hline
 craft\_rphp4\_085 & unsat & $177.09$ s \\
 \hline
 craft\_rphp4\_090 & unsat & $118.25$ s\\
 \hline
 craft\_rphp4\_095 & unsat & $197.94$ s\\
 \hline
 craft\_rphp4\_100 & unsat & $217.43$s\\
 \hline
 craft\_rphp4\_105 & unsat & $401.78$ s \\
 \hline
 craft\_rphp4\_110 & unsat & $698.25$ s \\
 \hline
 craft\_rphp4\_115 & unsat & $865.15$ s \\
 \hline
 craft\_rphp4\_120 & unsat & $1296.42$ s \\
 \hline
 craft\_rphp4\_125 & unsat & $1280.32$ s \\
 \hline
 craft\_rphp4\_130 & unsat & $1690.78$ s \\
  \hline
 craft\_rphp4\_135 & unsat & $1530.62$ s \\
  \hline
 craft\_rphp4\_140 & unsat & $2562.48$ s \\
  \hline
 craft\_rphp4\_145 & unsat & $1266.03$ s \\
  \hline
 craft\_rphp4\_150 & unsat & $2100.18$ s \\
  \hline
 craft\_rphp4\_155 & unsat & $3982.60$ s \\
  \hline
 craft\_rphp4\_160 & unsat & $4671.99$ s \\
 \hline
 craft\_rphp5\_035 & unsat & $69.89$ s \\
 \hline
 craft\_rphp5\_040 & unsat & $66.08$ s \\
 \hline
 craft\_rphp5\_045 & unsat & $253.78$ s \\
 \hline
 craft\_rphp5\_050 & unsat & $370.07$ s \\
 \hline
 craft\_rphp5\_055 & unsat & $355.40$ s\\
 \hline
 craft\_rphp5\_060 & unsat & $2122.91$ s\\
 \hline
 craft\_rphp5\_065 & unsat & $529.26$ s\\
 \hline
 craft\_rphp5\_070 & unsat & $928.81$ s \\
 \hline
 craft\_rphp5\_075 & unsat & $2529.35$ s \\
 \hline
 craft\_rphp5\_080 & unsat & $528.06$ s \\
 \hline
 craft\_rphp5\_085 & unsat & $717.96$ s \\
 \hline
 craft\_rphp5\_090 & unsat & $1368.22$ s \\
 \hline
 craft\_rphp5\_095 & unsat & $2429.29$ s \\
 \hline
 craft\_rphp5\_100 & unsat & $1018.40$ s \\
 \hline
 craft\_rphp5\_105 & unsat & $3916.32$ s \\
 \hline
 craft\_Ptn-7824-b01\ & unsat & $180.31$ s \\
 \hline
  craft\_Ptn-7824-b02\ & unsat & $173.49$ s \\
 \hline
  craft\_Ptn-7824-b03\ & unsat & $169.88$ s \\
 \hline
 craft\_Ptn-7824-b03\ & unsat & $169.88$ s \\
 \hline
  craft\_Ptn-7824-b04\ & unsat & $170.35$ s \\
 \hline
  craft\_Ptn-7824-b05\ & unsat & $167.99$ s \\
 \hline
  craft\_Ptn-7824-b06\ & unsat & $164.85$ s \\
  \hline
  craft\_Ptn-7824-b07\ & unsat & $451.97$ s \\
  \hline
  craft\_Ptn-7824-b08\ & unsat & $434.41$ s \\
  \hline
  craft\_Ptn-7824-b09\ & unsat & $438.03$ s \\
  \hline
  craft\_Ptn-7824-b010\ & unsat & $193.88$ s \\
 \hline
\end{supertabular}
\end{center}

 From Table \ref{Result-Craft}, it is clear that Gurobi solver based on 0-1 Linear Inequalities Model performances well. In SAT competition 2016, the best solver tc\_glucose for main-craft totally solves 58 instances used servers with good configuration \cite{Crafted}.
\section{Discussion and Conclusion}
In this paper, we presented 0-1 ILP. A key idea is to reduce the size of SSP matrix from $2(n+k)(n+k)$ to $n(n+k)$ using the property of this matrix, i.e., lines $v_i $and $v_i'$ located in can not be chosen in the same time and slack variables are all in the lower right corner.

Being different from reduction in \cite{Cormen2005Introduction}, LIMSAT works for general SAT problems including 3SAT but dos not need transforming SAT to 3SAT. We can construct a new SSP matrix according the CNF file. Comparing to the original SSP matrix introduced in \cite{Cormen2005Introduction}, the new one only contains variables and clauses without slack variables.

The experimental results show that our algorithm is effective. For future work, we will improve the efficiency of our implementation will be improved. Specifically, we consider using parallel algorithm to deal with SAT problems.

% conference papers do not normally have an appendix

% use section* for acknowledgment
%\section*{Acknowledgment}

% trigger a \newpage just before the given reference
% number - used to balance the columns on the last page
% adjust value as needed - may need to be readjusted if
% the document is modified later
%\IEEEtriggeratref{8}
% The "triggered" command can be changed if desired:
%\IEEEtriggercmd{\enlargethispage{-5in}}

% references section

% can use a bibliography generated by BibTeX as a .bbl file
% BibTeX documentation can be easily obtained at:
% http://mirror.ctan.org/biblio/bibtex/contrib/doc/
% The IEEEtran BibTeX style support page is at:
% http://www.michaelshell.org/tex/ieeetran/bibtex/
%\bibliographystyle{IEEEtran}
% argument is your BibTeX string definitions and bibliography database(s)
%\bibliography{IEEEabrv,../bib/paper}
%
% <OR> manually copy in the resultant .bbl file
% set second argument of \begin to the number of references
% (used to reserve space for the reference number labels box)

% that's all folks
\end{document}